\documentclass[aps,prd,preprint,superscriptaddress]{revtex4-1}
\usepackage{amsmath}
\usepackage{latexsym}
\usepackage{amsfonts}
\usepackage{graphicx}
\usepackage{mathrsfs}
\usepackage{epstopdf}
\usepackage{subfigure}
\usepackage{CJK}
\usepackage{hyperref}

\begin{document}

\title{Reconstruction of constant slow-roll inflation}

\author{Qing Gao}
\email{gaoqing1024@swu.edu.cn}
\affiliation{School of Physical Science and Technology, Southwest University, Chongqing 400715, China}


\begin{abstract}
By using the relations between the slow-roll parameters and the power spectrum for the single field slow-roll inflation, we derive the scalar spectral tilt $n_s$ and the
tensor to scalar ratio $r$ for the constant slow-roll inflation and obtain the constraint on the slow-roll parameter $\eta$ from
the Planck 2015 results. The inflationary potential for the constant slow-roll inflation is then reconstructed
in the framework of both general relativity and scalar-tensor theory of gravity, and compared with the recently reconstructed E model potential.
In the strong coupling limit, we show that the $\eta$ attractor is reached.
\end{abstract}

\preprint{1704.08559}

\maketitle
\section{Introduction}

The observational result on the scalar spectral tilt $n_s=0.9645\pm 0.0049$ (68\% CL) \cite{Adam:2015rua, Ade:2015lrj}
implies that $n_s-1\approx -2/N$, where $N$ is the number of $e$-folds before the end of inflation and we choose $N=60$.
So it is natural to parameterize observables $n_s$ and $r$ with $N$.
The parametrization of the slow-roll parameter $\epsilon$ by $N$ \cite{Huang:2007qz} was used to discuss the observables $n_s$ and $r$
and the sub-Planckian field excursion \cite{Lyth:1996im, Gong:2014cqa, Gao:2014yra, Gao:2014pca, Huang:2015xda,Linde:2016hbb}.
Mukhanov used the simple power-law parametrization $\epsilon(N)=\beta/(N+1)^\alpha$ to reconstruct the class of inflationary potentials \cite{Mukhanov:2013tua}.
The reconstruction of inflationary potentials were then discussed by
many researchers \cite{Roest:2013fha,Garcia-Bellido:2014eva, Garcia-Bellido:2014wfa, Garcia-Bellido:2014gna, Boubekeur:2014xva, Creminelli:2014nqa, Barranco:2014ira, Gobbetti:2015cya, Chiba:2015zpa,Binetruy:2014zya,Pieroni:2015cma,Binetruy:2016hna,Cicciarella:2016dnv,Barbosa-Cendejas:2015rba,Lin:2015fqa,Yi:2016jqr,Gao:2017uja,Odintsov:2017qpp,Nojiri:2017qvx}.
In this reconstruction method, the observables $n_s$ and $r$ are derived straightforwardly once we specify the parametrization
and the parameters can be constrained from observational data even before we derive the potentials \cite{Lin:2015fqa}.
Furthermore, the class of potentials are reconstructed in the full form, not just the first few terms in
Taylor expansion \cite{Hodges:1990bf,Copeland:1993jj, Liddle:1994cr, Lidsey:1995np,Ma:2014vua, Peiris:2006ug, Norena:2012rs, Choudhury:2014kma}.
Since the parametrization in terms of $N$ works on the observable scale only,
so there are some shortcomings on the reconstruction method \cite{Garcia-Bellido:2014gna,Martin:2016iqo}.

On the other hand, the attractor $n_s=1-2/N$ and $r=12/N^2$ can be derived from the T model \cite{Kallosh:2013hoa}, E model \cite{Kallosh:2013maa},
the Higgs inflation with the nonminimal coupling $\xi\psi^2 R$ in the
strong coupling limit $\xi\gg 1$ \cite{Kaiser:1994vs,Bezrukov:2007ep}, the more general potential
$\lambda^2 f^2(\psi)$ with the nonminimal coupling $\xi f(\psi)R$ for
arbitrary functions $f(\psi)$ in the strong coupling limit \cite{Kallosh:2013tua}, and the Starobinsky model $R+R^2$ \cite{starobinskyfr}.
The above attractor was also generalized to the so called $\alpha$ attractor with $n_s=1-2/N$ and $r=12\alpha/N^2$ \cite{Kallosh:2013yoa,Brooker:2017vyi,Jinno:2017jxc}.
Due to the arbitrary nonminimal coupling $\Omega(\phi)=1+\xi f(\phi)$ and the conformal transformation between Joran frame and Einstein frame,
the general scalar-tensor theories of gravity in Jordan frame can be brought into Einstein gravity plus canonical scalar field minimally coupled to gravity
in Einstein frame. Therefore,
in general, it is possible to obtain any attractor from general scalar-tensor theories of gravity \cite{Galante:2014ifa, Yi:2016jqr}.

In this paper, we discuss the constant slow-roll inflationary model \cite{Martin:2012pe,Motohashi:2014ppa,Motohashi:2017aob}.
If the slow-roll parameter $\epsilon$ is constant, then the other slow-roll parameter $\eta$ is 0 and inflation won't never stop.
So the constant slow-roll inflationary model means that $\eta$ is a constant,
here $\eta$ can be either the slow-roll parameter defined by the Hubble parameter or the potential.
We reconstruct the class of inflationary potentials for the constant slow-roll inflation in the framework of both general relativity and scalar-tensor theory of gravity.
We also fit the parameter $\eta$ to the observational data given by Planck observations \cite{Adam:2015rua, Ade:2015lrj}. The paper is organized as follows. In Sec. II,
we give the general formula and procedure for the reconstruction of the potentials with constant $\eta$ and compare
the potential with the reconstructed E model potential in \cite{DiMarco:2017sqo}. We conclude the paper in Sec. III.

\section{The constant slow-roll inflationary model}

For the constant slow-roll inflationary model \cite{Martin:2012pe,Motohashi:2014ppa,Motohashi:2017aob},
$\eta$ is a constant and $|\eta|<1$,
\begin{equation}
\label{etaeq1}
\eta=\frac{1}{V}\frac{d^2V}{d\phi^2},
\end{equation}
where the reduced Planck mass $M_{pl}=\sqrt{1/(8\pi G)}=1$. It is easy to see that the potential takes the form
\begin{equation}
\label{etaeq2}
V(\phi)=\begin{cases}
Ae^{\sqrt{\eta}\phi}+Be^{-\sqrt{\eta}\phi}, & 1>\eta>0\\
A+B\phi, & \eta=0\\
A\cos(\sqrt{-\eta}\,\phi)+B\sin(\sqrt{-\eta}\,\phi), & -1<\eta<0.
\end{cases}
\end{equation}
In the following, we use the reconstruction method to determine the integration constants $A$ and $B$.

\begin{figure*}[htbp]
$\begin{array}{c}
\includegraphics[width=0.46\textwidth]{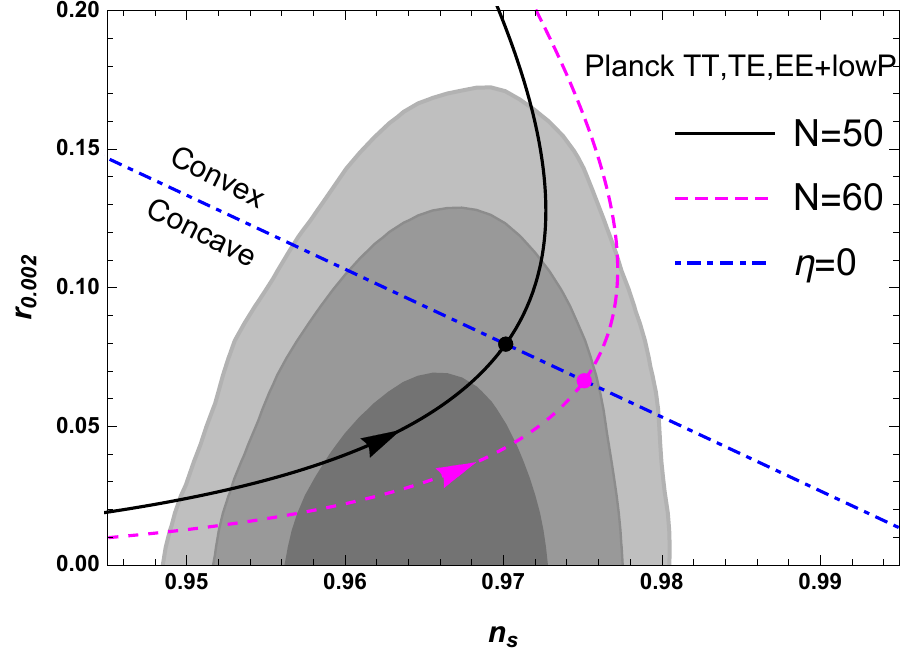}
\end{array}$
\caption{The marginalized 68\%, 95\% and 99.8\% confidence level contours for $n_s$ and $r_{0.002}$ from Planck 2015 data and the observational constraints on the constant slow-roll inflationary model. The black solid and magenta dashed lines correspond to $N=50$ and $N=60$, respectively, $\eta$ increases along the arrow direction, and the big dots
denote the values of $n_s$ and $r$ when $\eta=0$.}
\label{plketa}
\end{figure*}

From the relation
\begin{equation}
\label{nsapproxeq3}
2\eta=\frac{d\ln\epsilon}{dN}+4\epsilon,
\end{equation}
we get the solution
\begin{equation}
\label{epsilonn}
\epsilon(N)=\frac{\eta e^{2\eta N}}{D\eta+2e^{2\eta N}},
\end{equation}
where $D$ is an integration constant. In this paper, we assume the constant parametrization is valid during the whole inflation.
At the end of inflation, $N=0$, $\epsilon(N=0)\approx 1$, so $D=1-2/\eta$. The
slow-roll parameter \eqref{epsilonn} becomes
\begin{equation}
\label{epsilonn1}
\epsilon(N)=\frac{\eta e^{2\eta N}}{\eta-2+2e^{2\eta N}},
\end{equation}
and it has only one parameter $\eta$. Note that the sign of the denominator $\eta-2+2e^{2\eta N}$ in Eq. \eqref{epsilonn1} is the same as that of $\eta$.
The tensor to scalar ratio $r$ is
\begin{equation}
\label{rapprox}
r=16\epsilon=\frac{16\eta e^{2\eta N}}{\eta-2+2e^{2\eta N}},
\end{equation}
and the scalar spectral tilt $n_s$ is
\begin{equation}
\label{nsapproxeq4}
n_s=1+2\eta-6\epsilon=1+\frac{2\eta(\eta-2-e^{2\eta N})}{\eta-2+2e^{2\eta N}}.
\end{equation}
To get the constraint on the slow-roll parameter $\eta$,
we compare the results obtained from Eqs. \eqref{rapprox} and \eqref{nsapproxeq4} with the Planck 2015 observations \cite{Ade:2015lrj},
and the results are shown in Fig. \ref{plketa}. For $N=60$, the $1\sigma$ constraint is $-0.018<\eta<-0.0067$,
the $2\sigma$ constraint is $-0.021<\eta<0.0015$, and the $3\sigma$ constraint is $-0.023<\eta<0.01$.
For $N=50$, the $1\sigma$ constraint is $-0.014<\eta<-0.0039$, the $2\sigma$ constraint is $-0.018<\eta<0.0068$,
and the $3\sigma$ constraint is $-0.02<\eta<0.0168$. Since the denominator in Eq. \eqref{epsilonn1} becomes zero when $\eta=0$,
i.e., $\eta=0$ is a singular point, so we take Taylor
expansion of $\epsilon(N)$ around $\eta=0$,
\begin{equation}
\label{taylorepsilon}
\epsilon\approx\frac{1}{1+4N}.
\end{equation}
Plugging the result \eqref{taylorepsilon} into Eqs. \eqref{rapprox} and \eqref{nsapproxeq4}, we get
\begin{equation}
\label{taylornsr}
\begin{split}
 n_s&\approx 1-\frac{6}{1+4N},\\
 r&\approx \frac{16}{1+4N}.
 \end{split}
\end{equation}
When $\eta=0$, $(n_s, r)$ equal to $(0.97, 0.08)$ for $N=50$ and $(0.975, 0.067)$ for $N=60$, respectively.

From the definition of the slow-roll parameter $\epsilon$, we get
\begin{equation}
\label{dphidn1}
d\phi=\frac{dV/d\phi}{V}dN=\mp \sqrt{2\epsilon}dN,
\end{equation}
where the sign $\pm$ depends on the sign of the first derivative of the potential and the scalar
field is normalized by the reduced Planck mass $M_{pl}=1$. So
\begin{equation}
\label{nphieq2}
\phi-\phi_e=\pm \int_0^N \sqrt{2\epsilon(N)}dN,
\end{equation}
Substituting Eq. \eqref{epsilonn1} into
Eq. \eqref{nphieq2}, we get
\begin{equation}
\label{phineq2}
\phi=
\begin{cases}
\frac{\displaystyle 1}{\displaystyle \sqrt{\eta}}{\rm arctanh} \left(\sqrt{1-(2-\eta)e^{-2\eta N}/2}\,\right), & \eta>0,\\
\frac{\displaystyle 1}{\displaystyle \sqrt{-\eta}}\arctan\left(\sqrt{-1+(2-\eta)e^{-2\eta N}/2}\,\right),& \eta<0.
\end{cases}
\end{equation}
and the value $\phi_e$ of the scalar field at the end of inflation
\begin{equation}
\label{phineqe}
\phi_e=
\begin{cases}
\frac{\displaystyle 1}{\displaystyle \sqrt{\eta}}{\rm arctanh}\left(\sqrt{\eta/2}\right), & \eta>0,\\
\frac{\displaystyle 1}{\displaystyle \sqrt{-\eta}}\arctan\left(\sqrt{-\eta/2}\right), & \eta<0.
\end{cases}
\end{equation}
From the definition of the slow-roll parameter and the relation \eqref{dphidn1}, we get \cite{Chiba:2015zpa,Lin:2015fqa}
\begin{equation}
\label{sclreq2}
\epsilon=\frac{1}{2}\frac{dV/d\phi}{V}\frac{d\phi}{dN}=\frac{1}{2}\frac{d\ln V}{dN}.
\end{equation}
Plugging Eq. \eqref{epsilonn1} into Eq. \eqref{sclreq2}, we get
\begin{equation}
\label{vneq1}
V(N)=\tilde V_0\sqrt{|2-\eta-2e^{2\eta N}|}.
\end{equation}
Combining Eqs. \eqref{phineq2} and \eqref{vneq1}, we get the reconstructed potential in general relativity
\begin{equation}
\label{vphieq1}
V(\phi)=V_0\sin({\sqrt{-\eta}\phi}),
\end{equation}
where $V_0=\tilde V_0/\sqrt{-\eta}$. Note that if $\eta>0$, then the function $\sin$ becomes the function $\sinh$.

From Fig. \ref{plketa}, we find that $\eta>0$ is inconsistent with the
observations at the $1\sigma$ level, so in the following we consider $\eta<0$ only.
From Eqs. \eqref{phineq2} and \eqref{phineqe}, we get the field excursion
\begin{equation}
\label{alplbd}
\Delta\phi=\phi_*-\phi_e=\frac{1}{\sqrt{-\eta}}\left[\arctan\left(\sqrt{-1+\frac{2-\eta}{2}e^{-2\eta N}}\right)-\arctan\left(\sqrt{\frac{-\eta}{2}}\right) \right].
\end{equation}
From Eq. \eqref{alplbd}, if we take $\eta=-0.015$ and $N=60$, we get the super-Planckian field excursion $\Delta\phi=8.715$
which is bigger than the Lyth bound \cite{Lyth:1996im} $N\sqrt{r/8}=3.256$.

\begin{figure*}[htbp]
$\begin{array}{c}
\includegraphics[width=0.46\textwidth]{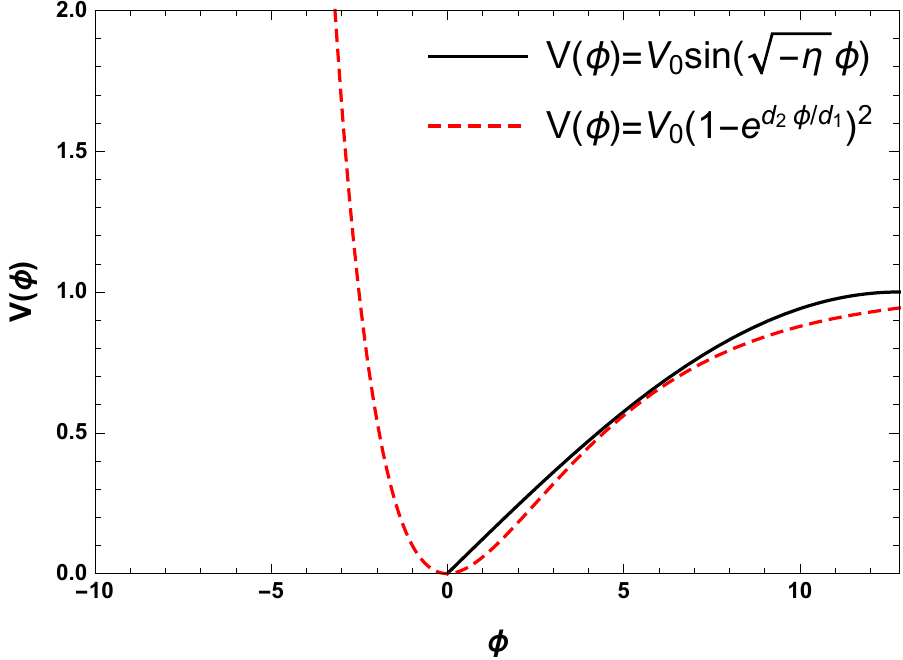}
\end{array}$
\caption{The comparison between different reconstructed potentials with $\eta=-0.015$.
The black solid line denotes our reconstructed potential \eqref{vphieq1} and the red dashed line denotes the reconstructed
E model potential \eqref{vphieq2} in \cite{DiMarco:2017sqo}. }
\label{fig2}
\end{figure*}

Now we compare the potential $V(\phi)=V_0\sin({\sqrt{-\eta}\phi})$ with the reconstructed E model potential \cite{DiMarco:2017sqo}
\begin{equation}
\label{vphieq2}
V(\phi)=V_0\left(1-e^{d_2\phi/d_1}\right)^2,
\end{equation}
where
\begin{equation}
\label{vphieq3}
d_1=\frac{1}{2}\sqrt{\frac{r}{2}},\ d_2=\frac{1}{3}\left[\frac{9r}{16}-\frac{3}{2}(1-n_s)\right].
\end{equation}
Take $\eta=-0.015$, we show the reconstructed potentials \eqref{vphieq1} and \eqref{vphieq2} in Fig. \ref{fig2}.
It is clear that the reconstructed potentials are consistent in the observable scale.

\begin{figure*}[htbp]
$\begin{array}{c}
\includegraphics[width=0.46\textwidth]{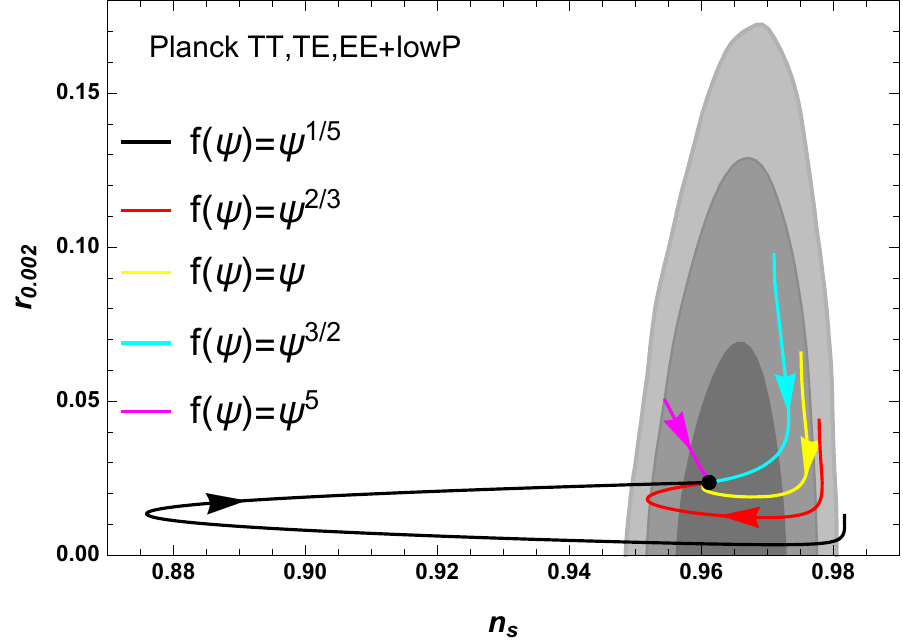}
\end{array}$
\caption{The numerical results of $n_s$ and $r$ for the reconstructed potential \eqref{eq51}.
We take $\omega(\psi)=1$, $N=60$, $\eta=-0.015$ and $f(\psi)=\psi^{k}$ with $k=1/5$, 2/3, 1, 3/2 and 5.
The coupling constant $\xi$ increases along the arrow direction.
The $\eta$ attractor is reached in the strong coupling limit. }
\label{fig3}
\end{figure*}

\begin{figure*}[htbp]
$\begin{array}{c}
\includegraphics[width=0.46\textwidth]{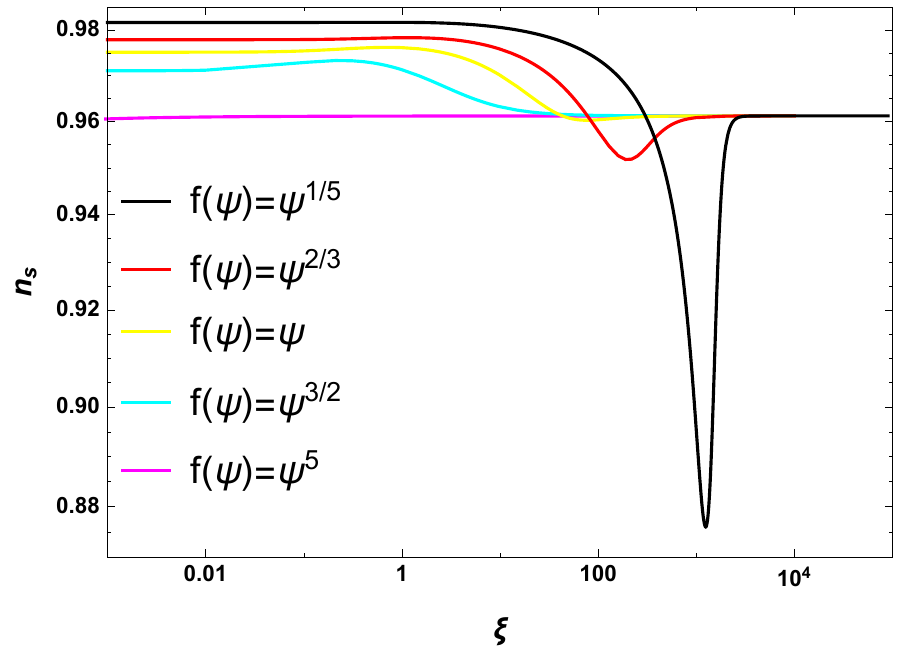}
\end{array}$
\caption{The dependence of $n_s$ on the coupling constant $\xi$. }
\label{fig4}
\end{figure*}

\begin{figure*}[htbp]
$\begin{array}{c}
\includegraphics[width=0.46\textwidth]{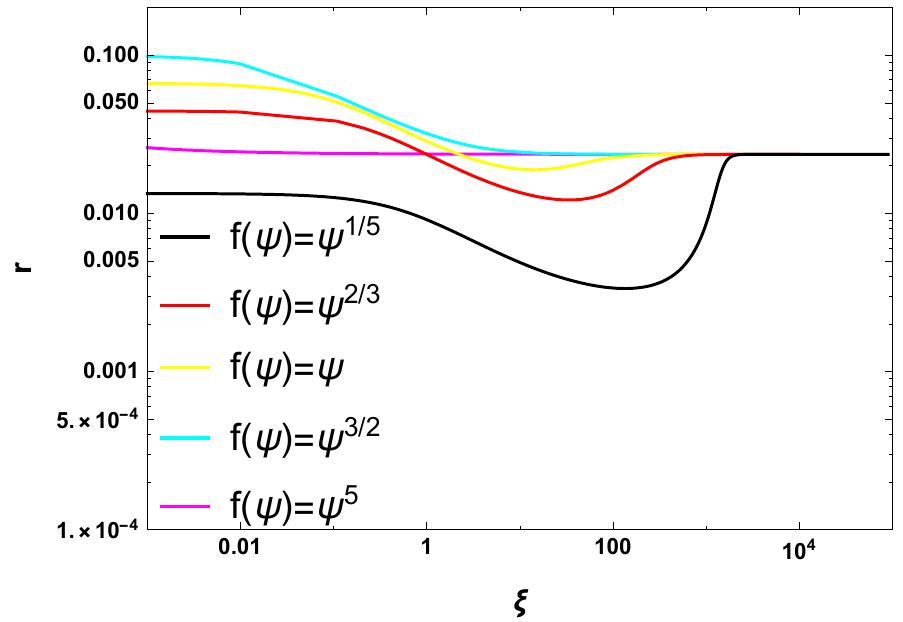}
\end{array}$
\caption{The dependence of $r$ on the coupling constant $\xi$. }
\label{fig5}
\end{figure*}

For the scalar field minimally coupled to gravity in Einstein frame, after the
the conformal transformation between the metric $\tilde{g}_{\mu\nu}$ and the scalar field $\psi$ in Jordan frame and the metric
$g_{\mu\nu}$ and the scalar field $\phi$ in Einstein frame,
\begin{gather}
\label{conftransf1}
g_{\mu\nu}=\Omega(\psi){\tilde g}_{\mu\nu},\\
\label{conftransf2}
d\phi^2=\left[\frac{3}{2}\frac{(d\Omega/d\psi)^2}{\Omega^2(\psi)}+\frac{\omega(\psi)}{\Omega(\psi)}\right]d\psi^2,
\end{gather}
we get the action for scalar-tensor theory in Jordan frame
\begin{equation}
\label{jscten} S=\int d^4x\sqrt{-\tilde{g}}\left[\frac{1}{2}\Omega(\psi)\tilde{R}(\tilde{g})-{1\over
2}\omega(\psi)\tilde{g}^{\mu\nu}
\nabla_{\mu}\psi\nabla_{\nu}\psi-V_J(\psi)\right],
\end{equation}
where $V_J(\psi)=\Omega^2(\psi)V(\phi)$. If the conformal factor satisfies the strong coupling condition
\begin{equation}
\label{strongcoup1}
\Omega(\psi)\ll \frac{3(d\Omega(\psi)/d\psi)^2}{2\omega(\psi)},
\end{equation}
then we get
\begin{equation}
\label{psiphirel1}
\phi\approx \sqrt{\frac{3}{2}}\ln\Omega(\psi),\quad \Omega(\psi)\approx e^{\sqrt{2/3}\,\phi}.
\end{equation}
For simplicity, we take $\Omega(\psi)=1+\xi f(\psi)$ and use the
above approximate relations \eqref{psiphirel1} in the strong coupling limit
to reconstruct the potential $V_J[\psi(\phi)]$ in Jordan frame.
For this specific
choice of $\Omega(\psi)$ with $f(\psi)=\psi^k$,
the strong coupling conditions \eqref{strongcoup1} and \eqref{psiphirel1}
become \cite{Gao:2017uja}
\begin{equation}
\label{strongcoup2}
\xi\gg \left(\frac{2}{3k^2}\right)^{k/2}\left(e^{\sqrt{2/3}\,\phi}-1\right)^{1-k}\exp\left(\sqrt{\frac{1}{6}}\,k\phi\right).
\end{equation}
Therefore, in the strong coupling limit, we get the reconstructed potential of the constant slow-roll inflation in the framework of scalar-tensor
theory of gravity,
\begin{equation}
\label{eq51}
V_J(\psi)=V_0\Omega^2(\psi)\sin\left(\sqrt{-\frac{3\eta}{2}}\ln\Omega(\psi)\right).
\end{equation}
Note that the function $\Omega(\psi)=1+\xi f(\psi)$ is arbitrary, so
we obtain the constant slow-roll inflationary attractor \eqref{rapprox} and \eqref{nsapproxeq4}
from the above potential \eqref{eq51} in the strong coupling limit \eqref{strongcoup1}, we call this attractor the $\eta$ attractor.
In Fig. \ref{fig3}, we take $\omega(\psi)=1$, $N=60$, $\eta=-0.015$ and $f(\psi)=\psi^{k}$ with $k=1/5$, 2/3, 1, 3/2 and 5
as examples to show the $\eta$ attractor $n_s=0.961$ and $r=0.024$ in the strong coupling limit. From Eq. \eqref{strongcoup2}, we find
that the strong coupling limit requires $\xi\gg 1933$ for $k=1/5$ and $\xi\gg 0.0013$ for $k=5$, the
dependences of $n_s$ and $r$ on the coupling constant $\xi$ are shown in
Figs. \ref{fig4} and \ref{fig5}. The results confirm the strong coupling condition \eqref{strongcoup2}.

\section{Conclusions}
We use the relations between observables and slow-roll parameters for the single field slow-roll inflation
to reconstruct the inflationary potential in the framework of both general relativity and scalar-tensor theory of gravity
by assuming that the slow-roll parameter $\eta$ is a constant. For the constant slow-roll inflation with constant $\eta$,
we first derive $\epsilon(N)$ from the relation between $\epsilon$ and $\eta$,
then we get the observalbes $r(N)$ and $n_s(N)$.
We compare the theoretical predications with the Planck 2015 observations \cite{Ade:2015lrj},
and the results are shown in Fig. \ref{plketa}. For $N=60$, the $1\sigma$ constraint is $-0.018<\eta<-0.0067$,
the $2\sigma$ constraint is $-0.021<\eta<0.0015$, and the $3\sigma$ constraint is $-0.023<\eta<0.01$.
For $N=50$, the $1\sigma$ constraint is $-0.014<\eta<-0.0039$, the $2\sigma$ constraint is $-0.018<\eta<0.0068$,
and the $3\sigma$ constraint is $-0.02<\eta<0.0168$. These results show that $\eta>0$ is inconsistent with the
observations at the $1\sigma$ level, so the observation favors $\eta<0$ and the concave potential $V(\phi)$ at the $1\sigma$ level.
From the relations between $\phi(N)$ and $\epsilon(N)$, $V(N)$ and $\epsilon(N)$, we get the
reconstructed potential $V(\phi)=V_0\sin({\sqrt{-\eta}\phi})$, and the result was compared
with the reconstructed E model potential in \cite{DiMarco:2017sqo}.
We find that the reconstructed potentials are consistent with each other in the observable scale.
Finally, we use the conformal transformation between Jordan frame and Einstein to reconstruct the class of extended
inflationary potentials, and the $\eta$ attractor is reached in the strong coupling limit as shown in Fig. \ref{fig3}.
We also use the strong coupling condition Eqs. \eqref{strongcoup2} to derive the constraint on the coupling constant $\xi$.
The derived analytical results are supported by the numerical results as shown in Figs. \ref{fig4} and \ref{fig5}.

\begin{acknowledgments}
The author thanks Professor Yungui Gong for helpful discussions. This research was supported in part by the National Natural Science
Foundation of China under Grant No. 11605061 and the Fundamental Research Funds for the
Central Universities.
\end{acknowledgments}


\begin{thebibliography}{10}

\bibitem{Adam:2015rua}
Adam R, et~al. {Planck 2015 results. I. Overview of products and scientific
  results}. Astron Astrophys, 2016, 594: A1

\bibitem{Ade:2015lrj}
Ade P~A~R, et~al. {Planck 2015 results. XX. Constraints on inflation}. Astron
  Astrophys, 2016, 594: A20

\bibitem{Huang:2007qz}
Huang Q~G. {Constraints on the spectral index for the inflation models in
  string landscape}. Phys Rev D, 2007, 76: 061303

\bibitem{Lyth:1996im}
Lyth D~H. {What would we learn by detecting a gravitational wave signal in the
  cosmic microwave background anisotropy?} Phys Rev Lett, 1997, 78: 1861--1863

\bibitem{Gong:2014cqa}
Gao Q, Gong Y. {The challenge for single field inflation with BICEP2 result}.
  Phys Lett B, 2014, 734: 41--43

\bibitem{Gao:2014yra}
Gao Q, Gong Y, Li T, et~al. {Simple single field inflation models and the
  running of spectral index}. Sci China Phys Mech Astron, 2014, 57: 1442--1448

\bibitem{Gao:2014pca}
Gao Q, Gong Y, Li T. {Modified Lyth bound and implications of BICEP2 results}.
  Phys Rev D, 2015, 91(6): 063509

\bibitem{Huang:2015xda}
Huang Q~G. {Lyth bound revisited}. Phys Rev D, 2015, 91(12): 123532

\bibitem{Linde:2016hbb}
Linde A. {Gravitational waves and large field inflation}. JCAP, 2017, 1702(02):
  006

\bibitem{Mukhanov:2013tua}
Mukhanov V. {Quantum Cosmological Perturbations: Predictions and Observations}.
  Eur Phys J C, 2013, 73: 2486

\bibitem{Roest:2013fha}
Roest D. {Universality classes of inflation}. JCAP, 2014, 1401(01): 007

\bibitem{Garcia-Bellido:2014eva}
Garcia-Bellido J, Roest D, Scalisi M, et~al. {Can CMB data constrain the
  inflationary field range?} JCAP, 2014, 1409: 006

\bibitem{Garcia-Bellido:2014wfa}
Garcia-Bellido J, Roest D, Scalisi M, et~al. {Lyth bound of inflation with a
  tilt}. Phys Rev D, 2014, 90(12): 123539

\bibitem{Garcia-Bellido:2014gna}
Garcia-Bellido J, Roest D. {Large-$N$ running of the spectral index of
  inflation}. Phys Rev D, 2014, 89(10): 103527

\bibitem{Boubekeur:2014xva}
Boubekeur L, Giusarma E, Mena O, et~al. {Phenomenological approaches of
  inflation and their equivalence}. Phys Rev D, 2015, 91(8): 083006

\bibitem{Creminelli:2014nqa}
Creminelli P, Dubovsky S, López~Nacir D, et~al. {Implications of the scalar
  tilt for the tensor-to-scalar ratio}. Phys Rev D, 2015, 92(12): 123528

\bibitem{Barranco:2014ira}
Barranco L, Boubekeur L, Mena O. {A model-independent fit to Planck and BICEP2
  data}. Phys Rev D, 2014, 90(6): 063007

\bibitem{Gobbetti:2015cya}
Gobbetti R, Pajer E, Roest D. {On the Three Primordial Numbers}. JCAP, 2015,
  1509(09): 058

\bibitem{Chiba:2015zpa}
Chiba T. {Reconstructing the inflaton potential from the spectral index}. Prog
  Theor Exp Phys, 2015, 2015(7): 073E02

\bibitem{Binetruy:2014zya}
Binetruy P, Kiritsis E, Mabillard J, et~al. {Universality classes for models of
  inflation}. JCAP, 2015, 1504(04): 033

\bibitem{Pieroni:2015cma}
Pieroni M. {$\beta$-function formalism for inflationary models with a non
  minimal coupling with gravity}. JCAP, 2016, 1602(02): 012

\bibitem{Binetruy:2016hna}
Binétruy P, Mabillard J, Pieroni M. {Universality in generalized models of
  inflation}.
\newblock \href{http://arxiv.org/abs/1611.07019}{{ arXiv: 1611.07019 [gr-qc]}}

\bibitem{Cicciarella:2016dnv}
Cicciarella F, Pieroni M. {Universality for quintessence}.
\newblock \href{http://arxiv.org/abs/1611.10074}{{ arXiv: 1611.10074 [gr-qc]}}

\bibitem{Barbosa-Cendejas:2015rba}
Barbosa-Cendejas N, De-Santiago J, German G, et~al. {Tachyon inflation in the
  Large-$N$ formalism}. JCAP, 2015, 1511: 020

\bibitem{Lin:2015fqa}
Lin J, Gao Q, Gong Y. {The model independent reconstruction of inflationary
  potentials}. Mon Not Roy Astron Soc, 2016, 459: 4029--4037

\bibitem{Yi:2016jqr}
Yi Z, Gong Y. {Nonminimal coupling and inflationary attractors}. Phys Rev D,
  2016, 94(10): 103527

\bibitem{Gao:2017uja}
Gao Q, Gong Y. {Reconstruction of extended inflationary potentials for
  attractors}.
\newblock \href{http://arxiv.org/abs/1703.02220}{{ arXiv: 1703.02220 [gr-qc]}}

\bibitem{Odintsov:2017qpp}
Odintsov S~D, Oikonomou V~K. {Inflation with a Smooth Constant-Roll to
  Constant-Roll Era Transition}.
\newblock \href{http://arxiv.org/abs/1704.02931}{{ arXiv: 1704.02931 [gr-qc]}}

\bibitem{Nojiri:2017qvx}
Nojiri S, Odintsov S~D, Oikonomou V~K. {Constant-roll Inflation in $F(R)$
  Gravity}.
\newblock \href{http://arxiv.org/abs/1704.05945}{{ arXiv: 1704.05945 [gr-qc]}}

\bibitem{Hodges:1990bf}
Hodges H~M, Blumenthal G~R. {Arbitrariness of inflationary fluctuation
  spectra}. Phys Rev D, 1990, 42: 3329--3333

\bibitem{Copeland:1993jj}
Copeland E~J, Kolb E~W, Liddle A~R, et~al. {Reconstructing the inflation
  potential, in principle and in practice}. Phys Rev D, 1993, 48: 2529--2547

\bibitem{Liddle:1994cr}
Liddle A~R, Turner M~S. {Second order reconstruction of the inflationary
  potential}. Phys Rev D, 1994, 50: 758

\bibitem{Lidsey:1995np}
Lidsey J~E, Liddle A~R, Kolb E~W, et~al. {Reconstructing the inflation
  potential : An overview}. Rev Mod Phys, 1997, 69: 373--410

\bibitem{Ma:2014vua}
Ma Y~Z, Wang Y. {Reconstructing the Local Potential of Inflation with BICEP2
  data}. JCAP, 2014, 1409(09): 041

\bibitem{Peiris:2006ug}
Peiris H, Easther R. {Recovering the Inflationary Potential and Primordial
  Power Spectrum With a Slow Roll Prior: Methodology and Application to WMAP 3
  Year Data}. JCAP, 2006, 0607: 002

\bibitem{Norena:2012rs}
Norena J, Wagner C, Verde L, et~al. {Bayesian Analysis of Inflation III: Slow
  Roll Reconstruction Using Model Selection}. Phys Rev D, 2012, 86: 023505

\bibitem{Choudhury:2014kma}
Choudhury S, Mazumdar A. {Reconstructing inflationary potential from BICEP2 and
  running of tensor modes}.
\newblock \href{http://arxiv.org/abs/1403.5549}{{ arXiv: 1403.5549 [hep-th]}}

\bibitem{Martin:2016iqo}
Martin J, Ringeval C, Vennin V. {Shortcomings of New Parametrizations of
  Inflation}. Phys Rev D, 2016, 94(12): 123521

\bibitem{Kallosh:2013hoa}
Kallosh R, Linde A. {Universality Class in Conformal Inflation}. JCAP, 2013,
  1307: 002

\bibitem{Kallosh:2013maa}
Kallosh R, Linde A. {Non-minimal Inflationary Attractors}. JCAP, 2013, 1310:
  033

\bibitem{Kaiser:1994vs}
Kaiser D~I. {Primordial spectral indices from generalized Einstein theories}.
  Phys Rev D, 1995, 52: 4295--4306

\bibitem{Bezrukov:2007ep}
Bezrukov F~L, Shaposhnikov M. {The Standard Model Higgs boson as the inflaton}.
  Phys Lett B, 2008, 659: 703--706

\bibitem{Kallosh:2013tua}
Kallosh R, Linde A, Roest D. {A universal attractor for inflation at strong
  coupling}. Phys Rev Lett, 2014, 112: 011303

\bibitem{starobinskyfr}
Starobinsky A~A. {A New Type of Isotropic Cosmological Models Without
  Singularity}. Phys Lett B, 1980, 91: 99--102

\bibitem{Kallosh:2013yoa}
Kallosh R, Linde A, Roest D. {Superconformal Inflationary $\alpha$-Attractors}.
  JHEP, 2013, 1311: 198

\bibitem{Brooker:2017vyi}
Brooker D~J. {How to Produce an Arbitrarily Small Tensor to Scalar Ratio}.
\newblock \href{http://arxiv.org/abs/1703.07225}{{ arXiv: 1703.07225
  [astro-ph.CO]}}

\bibitem{Jinno:2017jxc}
Jinno R, Kaneta K. {Hillclimbing inflation}.
\newblock \href{http://arxiv.org/abs/1703.09020}{{ arXiv: 1703.09020 [hep-ph]}}

\bibitem{Galante:2014ifa}
Galante M, Kallosh R, Linde A, et~al. {Unity of Cosmological Inflation
  Attractors}. Phys Rev Lett, 2015, 114(14): 141302

\bibitem{Martin:2012pe}
Martin J, Motohashi H, Suyama T. {Ultra Slow-Roll Inflation and the
  non-Gaussianity Consistency Relation}. Phys Rev D, 2013, 87(2): 023514

\bibitem{Motohashi:2014ppa}
Motohashi H, Starobinsky A~A, Yokoyama J. {Inflation with a constant rate of
  roll}. JCAP, 2015, 1509(09): 018

\bibitem{Motohashi:2017aob}
Motohashi H, Starobinsky A~A. {Constant-roll inflation: confrontation with
  recent observational data}. Europhys Lett, 2017, 117(3): 39001

\bibitem{DiMarco:2017sqo}
Di~Marco A, Cabella P, Vittorio N. {Reconstruction of $\alpha$-attractor
  supergravity models of inflation}. Phys Rev D, 2017, 95(2): 023516

\end{thebibliography}

\end{document}